# Vertex dynamics in multi-soliton solutions of Kadomtsev-Petviashvili II equation

Yair Zarmi
Jacob Blaustein Institutes for Desert Research
Ben-Gurion University of the Negev
Midreshet Ben-Gurion, 84990 Israel

Abstract

A functional of the solution of the Kadomtsev-Petviashvili II equation maps multi-soliton solutions onto systems of vertices – structures that are localized around soliton junctions. A solution with one junction is mapped onto a single vertex, which emulates a free, spatially extended, particle. In solutions with several junctions, each junction is mapped onto a vertex. Moving in the *x-y* plane, the vertices collide, coalesce upon collision, and then split up. When well separated, they emulate free particles. Multi-soliton solutions, whose structure does not change under space-time inversion as $|t| \to \infty$, are mapped onto vertex systems that undergo elastic collisions. Solutions, whose structure does change, are mapped onto systems that undergo inelastic collisions. The inelastic vertex collisions generated from the infinite family of (*M*,1) solutions (*M* external solitons, (*M*–2) *Y*-shaped soliton junctions, $M \geq 4$) play a unique role: The only definition of vertex mass consistent with momentum conservation in these collisions is the spatial integral of the vertex profile. This definition ensures, in addition, that, in these collisions, the total mass and kinetic energy due to the motion in the *y*-direction are conserved. In general, the kinetic energy due to the motion in the *x*-direction is not conserved in these collisions.

# 1. Introduction

Soliton solutions of integrable nonlinear evolution equations in (1+1) dimensions are localized in position at a given time. In some applications to classical systems, this has allowed for the interpretation of a soliton as a spatially extended but localized particle with a finite mass:

$$m = \int_{-\infty}^{+\infty} u(t,x)\,dx < \infty \quad . \tag{1}$$

When the evolution equation is associated with a quantum-dynamical system, then the solution has been sometimes viewed as a candidate for a normalizable wave function, since

$$\int_{-\infty}^{+\infty} |u(t,x)|^2\,dx < \infty \quad . \tag{2}$$

In quantum-field theoretical applications, such a wave function may be used as a starting point for the construction of bound states of the theory.

Line-solitons in more than one space dimension cannot be used for the goals delineated above because they are not localized in space. This has led to an extensive search for integrable nonlinear evolution equations in higher dimensions that have spatially localized solutions. Prime examples in the classical arena are the Kadomtsev-Petviashvili I [1,2], Davey-Stewartson [3,4], Gardner [5,6] and Nizhnik-Veselov-Novikov [7-9] equations. In Quantum-Field Theory, the search for such equations began with the discovery of the 't Hooft-Polyakov monopole [10,11] - a spatially localized solution of the (1+3)-dimensional nonlinear Klein-Gordon equation.

This paper presents a different approach to the generation of localized structures that emulate spatially extended particles in more than one space dimension. These structures are *not* solutions of an integrable nonlinear evolution equation. Rather, they are images of *line-soliton* solutions of such an equation. A functional of the solution, which vanishes on a single soliton, maps multi-soliton solutions onto vertices - structures that are localized around soliton junctions.

Vertex maps in (1+1) dimensions have been exploited in the analysis of the perturbed KdV equation [47]. However, vertex dynamics is then trivial. The vertices are localized in the $x$–$t$ plane; in time, they evolve and decay. Interesting dynamics emerges in higher space dimensions. Vertices then move in space, emulating spatially extended particles. In the present work, the collision dynamics of vertices, which emulate non-relativistic particles, is unraveled in the case of the line-soliton solutions of the Kadomtsev-Petviashvili II (KP II) equation [1]:

$$\frac{\partial}{\partial x}\left(-4\frac{\partial u}{\partial t} + \frac{\partial^3 u}{\partial x^3} + 6u\frac{\partial u}{\partial x}\right) + 3\frac{\partial^2 u}{\partial y^2} = 0 \quad . \tag{3}$$

The literature on soliton solutions of Eq. (3) is very rich [12-46]. Still, as preparation for the discussion of vertex dynamics, Section 2 is dedicated to a review of pertinent solution properties and to the presentation of solutions, which provide examples for statements made in the paper.

The new results are presented in Sections 3-6. In Section 3, it is first shown that a differential polynomial, $R[u]$, exists, which, as a direct consequence of Eq. (3), vanishes on the single-soliton solution. $R[u]$ maps multi-soliton solutions onto a collection of vertices - structures, each localized around one soliton junction. Solutions with one soliton junction (e.g., the *Y*"- [12-14, 28, 31-41] and "*X*"- [15, 31-41] shaped solutions) are mapped onto one vertex. Single-junction solutions preserve their spatial structure as they propagate at a constant speed. Hence, so do their single-vertex maps. Thus, an isolated vertex emulates a free, spatially extended particle.

Most multi-soliton solutions form web patterns of soliton lines that intersect in junctions. A soliton junction occurs around a point in space and time, at which some of the phases involved in the construction of the solution coincide [31-44]. For example, a junction of three solitons (a *Y*-shaped segment) occurs around a point, at which three phases coincide. At finite times, two, or more, junctions may be close to one another and eventually coalesce. This happens around points, at which more than three phases coincide. Identification of such points plays an essential role in the determination of the temporal and spatial evolution of KP II tree-shaped soliton solutions [41-44]. This situation is reflected in the vertex map of a web. At some finite times, the distances amongst vertices become comparable to their spatial extent. Upon collision, they lose their identities, coalesce, and then split up. As $|t| \to \infty$, the web tends to a system of external single solitons that meet in distinct junctions, which are connected by internal single solitons. Vanishing on single solitons, the functional $R[u]$ then maps the solution onto a collection of well-separated vertices, each an image of one soliton junction. For this reason, vertex collisions are analyzed in this paper only for $|t| \to \infty$.

Section 4 discusses the kinematics of vertex collisions. The main results of the paper are:

1) The collision of the vertices is elastic when they are generated from a solution whose structure does not change under space-time inversion as $|t| \to \infty$. If the structure does change, the map describes an inelastic vertex collision.

2) Momentum conservation in the inelastic vertex collision generated from the (4,1) solution (discussed in Section 2.2.3.1) requires that, up to a multiplicative constant, the mass of a vertex map of a *Y*-shaped solution, must be the space integral of vertex profile.

3) With this mass definition, in the inelastic collisions described by the vertex maps of the infinite family of (*M*,1)-multi-soliton solutions, (*M* external solitons, (*M*–2) junctions, $M \geq 4$), the total vertex momentum, mass and kinetic energy of the motion in the *y*-direction are all conserved. In general, the kinetic energy of the motion in the *x*-direction is not conserved.

These statements are demonstrated through numerical examples of vertex maps of solutions that are constructed from four wave numbers.

In soliton dynamics, localized structures have often been regarded as models for spatially extended particles, with particle mass usually defined as the space integral of the amplitude of the structure. When solitons represent massive objects (e.g., a shallow-water surface wave), this definition is natural. In other cases, it is an intuitively appealing assumption. In the vertex collisions discussed here this definition is essential for the interpretation of vertices as spatially extended particles that undergo collisions, in which momentum is conserved.

Finally, there is an infinite hierarchy of differential polynomials that vanish on the single-soliton solution. (The existence of such a hierarchy has been discussed in detail in the case of the KdV equation [47].) However, only $R[u]$ generates localized structures, which obey the conservation laws cited above. This is discussed in Section 5. Section 6 discusses preliminary results concerning the equations that govern vertex dynamics. Section 7 presents concluding comments.

## 2. Review of line-soliton solutions of KP II equation
### 2.1 Construction and general properties

The line-soliton solutions of Eq. (3) are obtained through a Hirota transformation [23, 24]:

$$u(t,x,y) = 2\partial_x^2 \log\{f(t,x,y)\} \ . \tag{4}$$

The function $f(t,x,y)$ is given by

$$f(t,x,y) \equiv f(M,N;\vec{\xi}) =$$

$$\begin{cases} \sum_{i=1}^{M} \xi_M(i)\exp(\theta_i(t,x,y)) & N = 1 \\ \sum_{i=1}^{M} \xi_M(i)\exp\left(\sum_{j=1,j\neq i}^{M} \theta_j(t,x,y)\right) & N = M-1 \\ \sum_{1\leq i_1<\ldots<i_N\leq M} \xi_M(i_1,\ldots,i_N)\left(\prod_{1\leq j<l\leq N}(k_{i_l}-k_{i_j})\right)\exp\left(\sum_{j=1}^{N}\theta_{i_j}(t,x,y)\right) & 2\leq N\leq M-2 \end{cases}, \tag{5}$$

$$k_1 < k_2 < \ldots < k_M \ , \tag{6}$$

$$\theta_i(t,x,y) = -k_i x + k_i^2 y - k_i^3 t \ . \tag{7}$$

In Eq. (5), $M$ is the size of a set of wave numbers. Each set of $N$ wave numbers is one of the $\binom{M}{N}$ subsets of $\{k_1,\ldots,k_M\}$. In the following, $f(t,x,y)$ and $u(t,x,y)$ will be denoted, respectively, by $f(M,N;\vec{\xi})$ and $u(M,N;\vec{\xi})$.

### 2.1.1 Constraints on $\xi_M(i_1,\ldots, i_M)$

To exclude singular solutions of Eq. (3), one requires

$$\xi_M(i_1,\ldots,i_N) \geq 0 \ . \tag{8}$$

Apart from positivity, the coefficients, $\xi_M(i)$, with $N = 1$ and $N = M-1$, may assume arbitrary values. However, for $2 \leq N \leq M-2$, $\xi_M(i_1,\ldots,i_N)$ are constrained by the Plücker relations (see, E.g. [30]). For example, for $(M,N) = (4,2)$ one finds a single Plücker relation:

$$\xi_4(1,2)\xi_4(3,4) - \xi_4(1,3)\xi_4(2,4) + \xi_4(1,4)\xi_4(2,3) = 0 \ . \tag{9}$$

### 2.1.2 Space-time inversion and structure of solutions

The properties of the soliton solutions of Eq. (3) under space-time inversion have been discussed in the literature [31-44]. However, because of its importance for the characteristics of vertex collisions discussed in Sections 3 and 4, this topic is reviewed here in some detail.

The solution, $u(M,N;\vec{\xi})$, is also generated by Eq. (5) when $f(M,N;\vec{\xi})$ is replaced by

$$\tilde{f}(M,N;\vec{\xi}) = f(M,N;\vec{\xi}) \Big/ \left( \prod_{i=1}^{M} \exp\theta_i(t,x,y) \right) =$$

$$\begin{cases} \sum_{i=1}^{M} \xi_M(i)\exp\left(-\sum_{j=1,j\neq i}^{N} \theta_j(t,x,y)\right) & N = 1 \\ \sum_{i=1}^{M} \xi_M(i)\exp\left(-\theta_i(t,x,y)\right) & N = M-1 \\ \sum_{1\leq i_1<\ldots<i_N\leq M} \xi_M(i_1,\ldots,i_N)\left(\prod_{1\leq j<l\leq N}(k_{i_l} - k_{i_j})\right)\exp\left(-\sum_{j=1,i_j\neq i_1,\ldots,i_N}^{N} \theta_{i_j}(t,x,y)\right) & 2 \leq N \leq M-1 \end{cases} \tag{10}$$

Under space-time inversion, $\tilde{f}(M,N;\vec{\xi})$ is transformed into

$$\tilde{f}(M,N;\vec{\xi}) \xrightarrow[(t,x,y)\to(-t,-x,-y)]{} \tilde{\tilde{f}}(M,M-N;\vec{\xi}) \ . \tag{11}$$

Each monomial in $f(M, N; \vec{\xi})$ is a product of $N$ exponentials, with phases that depend on a subset of wave numbers, $(k_{i_1},...,k_{i_N})$. $\tilde{f}(M, M–N; \vec{\xi})$ is obtained by replacing a monomial in $f(M, N; \vec{\xi})$ by a monomial that contains a product of $(M - N)$ exponentials, with phases that depend on the set of $(M - N)$ wave numbers, which are the *complement* of $(k_{i_1},...,k_{i_N})$. Thus, $\tilde{f}(M, M–N; \vec{\xi})$ generates through Eq. (5) two solutions: The original solution, $u(M, N; \vec{\xi})$, computed at $(-t, -x, -y)$ and a new solution, $u(M, M–N; \vec{\xi})$ computed at $(t, x, y)$. Consequently, the structure of $u(M, M; \vec{\xi})$ at $(-t, -x, -y)$ coincides with that of $u(M, M–N; \vec{\xi})$ at $(t, x, y)$.

For $M \geq 4$, but $M \neq 2N$, the exponential terms in the original function, $f(M, N; \vec{\xi})$, and in the transformed function, $\tilde{f}(M, M–N; \vec{\xi})$ of Eq. (11), are different. As an example, consider

$$f(4,1;\vec{\xi}) = \xi_4(1)e^{\theta_1} + \xi_4(2)e^{\theta_2} + \xi_4(3)e^{\theta_3} + \xi_4(4)e^{\theta_4}. \tag{12}$$

The transformation of Eq. (11) yields

$$\tilde{f}(4,3;\vec{\xi}) = \xi_4(1)e^{\theta_2+\theta_3+\theta_4} + \xi_4(2)e^{\theta_1+\theta_3+\theta_4} + \xi_4(3)e^{\theta_1+\theta_2+\theta_4} + \xi_4(4)e^{\theta_1+\theta_2+\theta_3}. \tag{13}$$

Thus, the structure of $u(4,1;\vec{\xi})$ at $(-t, -x, -y)$ is the same as the structure of $u(4,3;\vec{\xi})$ at $(t,x,y)$. As in these two solutions the interplay amongst phases is different (see Appendix I), the structure of $u(4,1;\vec{\xi})$ at $(t,x,y)$ is expected, in general, to be different from its structure at $(-t, x, y)$.

In contrast, for solutions with $M = 2N$, the transformed function, $\tilde{f}(2N, N; \vec{\xi})$ of Eq. (11), contains the *same* sets of $N$ exponentials as the original, $f(2N,N;\vec{\xi})$, with the $\xi$-coefficients exchanging places in the sum. For example, in the case of $(M,N) = (4,2)$, the function $f(4,2;\vec{\xi})$ is given by:

$$\begin{aligned}f(4,2;\vec{\xi}) = &\xi_4(1,2)e^{\theta_1+\theta_2} + \xi_4(1,3)e^{\theta_1+\theta_3} + \xi_4(1,4)e^{\theta_1+\theta_4} \\ &+ \xi_4(2,3)e^{\theta_2+\theta_3} + \xi_4(2,4)e^{\theta_2+\theta_4} + \xi_4(3,4)e^{\theta_3+\theta_4}\end{aligned}. \tag{14}$$

Applying Eqs. (10) and (11) to Eq. (14) yields the following result:

$$\begin{aligned}\tilde{f}(4,2;\vec{\xi}) = &\xi_4(3,4)e^{\theta_1+\theta_2} + \xi_4(2,4)e^{\theta_1+\theta_3} + \xi_4(2,3)e^{\theta_1+\theta_4} \\ &+ \xi_4(1,4)e^{\theta_2+\theta_3} + \xi_4(1,3)e^{\theta_2+\theta_4} + \xi_4(1,2)e^{\theta_3+\theta_4}\end{aligned} \quad (15)$$

Except for a different set of constant phase shifts, the structure of $\tilde{f}(4,2;\vec{\xi})$ of Eq. (15) coincides with that of the original $f(4,2;\vec{\xi})$. As $|t| \to \infty$, the constant shifts cannot affect the structure of the web of soliton lines. Consequently, up to these constant shifts, the structure of $u(4,2;\vec{\xi})$ in the *x-y* plane at $t \to -\infty$ and at $t \to +\infty$ are mirror images of one another.

## 2.2 Specific KP II line-soliton solutions
### 2.2.1 Two wave numbers – the single-soliton solution
The single-soliton solution of Eq. (3) is generated by

$$f(t,x,y) = \xi_1 e^{\theta(t,x,y;k_1)} + \xi_2 e^{\theta(t,x,y;k_2)} \quad \left(\theta(t,x,y;k) = -kx + k^2 y - k^3 t\right) . \quad (16)$$

In Eq. (16), $\xi_1$ and $\xi_2$ are arbitrary positive coefficients. The solution itself has the form:

$$u(t,x,y) = \frac{2\left((k_2-k_1)/2\right)^2}{\left(\cosh\left[\left((k_2-k_1)/2\right)\left(x-(k_1+k_2)y+(k_1^2+k_1 k_2+k_2^2)t\right)+\ln\left(\sqrt{\xi_1/\xi_2}\right)\right]\right)^2} . \quad (17)$$

### 2.2.2 Solutions with one junction
#### 2.2.2.1 Three wave numbers – The *Y*-shaped solution [12-14, 28, 31-41]
All solutions with *M* = 3 in Eq. (5) have one junction. These are the Miles-resonance [12-14] or *Y*-shaped [28, 31-41] solutions. With three wave numbers, there are two solutions with (*M*,*N*) = (3,1) and (3,2), which are related by space-time inversion. Hence, only the (3,1) solution is analyzed. Eq.(5) then has the form:

$$f(t,x,y) = \xi_1 e^{\theta(t,x,y;k_1)} + \xi_2 e^{\theta(t,x,y;k_2)} + \xi_3 e^{\theta(t,x,y;k_3)} \quad (18)$$

The *Y*-shaped solution preserves its shape in time as it moves at a constant velocity [28]. Substituting Eq. (18) in Eq. (5), one finds that a moving frame exists, such that this solution obeys:

$$u(t, x+v_x t, y+v_y t) = u(t=0, x, y) . \quad (19)$$

The velocity vector is given by

$$v_x(k_1,k_2,k_3) = k_1 k_2 + k_1 k_3 + k_2 k_3 \quad , \quad v_y(k_1,k_2,k_3) = k_1 + k_2 + k_3 \quad . \tag{20}$$

Fig. 1 shows a *Y*-shaped solution. Away from the junction, each line tends to a single soliton. The two numbers adjacent to a soliton indicate the pair of wave numbers that participate in its construction in Eq. (17). Wave-number assignment to solitons is described in Appendix I. It has been employed in the construction of all the numerical examples studied in this paper.

**2.2.2.2 Four different wave numbers – The *X*-shaped solution [15, 31-41]**
For $M \geq 4$, solutions with one junction exist only for specific choices of the $\xi$-coefficients in Eq. (5), and with some constraints on the wave numbers. The *X*-shaped solution (an $(M,N) = (4,2)$ solution) provides an example for this statement. An *X*-shaped solution (a numerical example of which is displayed in Fig. 2) exists, for instance, for the following choice of the $\xi$ - coefficients:

$$\xi_4(1,2) = \xi_4(3,4) = 0 \qquad \xi_4(1,3) = \xi_4(1,4) = \xi_4(2,3) = \xi_4(2,4) = 1 \quad . \tag{21}$$

Eq. (19) holds also for the *X*-shaped solution. It propagates at a constant velocity, preserving its shape at all times. (In fluid-dynamical experiments, the velocity was observed to be roughly constant, and decrease slowly as energy dissipation occurred [15].) For the choice of Eq. (21), the velocity is given by:

$$\begin{aligned}
v_x &= k_1 k_2 + k_1 k_3 + k_1 k_4 + k_2 k_3 + k_2 k_4 + k_3 k_4 - \frac{k_3 k_4 (k_3 + k_4) - k_1 k_2 (k_1 + k_2)}{(k_3 + k_4) - (k_1 + k_2)} \\
v_y &= k_1 + k_2 + k_3 + k_4 - \frac{k_3 k_4 - k_1 k_2}{(k_3 + k_4) - (k_1 + k_2)}
\end{aligned} \tag{22}$$

The velocity becomes infinite (the *X*-shaped solution degenerates into two parallel solitons) when

$$k_3 + k_4 = k_1 + k_2 \quad . \tag{23}$$

### 2.2.3 Four wave numbers - solutions with more than one soliton junction

The examples presented in the following deal with solutions generated from four wave numbers. Except for specific choices of the $\xi$-coefficients and/or the wave numbers, Eq. (5) then generates three four-soliton processes: As $t \to \pm\infty$, two solutions have two junctions ($(M,N) = (4,1)$ and (4,3), which are related by space-time inversion) and one solution has four ($(M,N) = (4,2)$).

#### 2.2.3.1 Four wave numbers – two soliton junctions: "Riding" on Y-shaped junctions

This Section focuses on the $(M,N) = (4,1)$ solution, which depends on four arbitrary positive coefficients, $\xi_i$, $1 \leq i \leq 4$, and is given by inserting in Eq. (4) the following function:

$$f_{\bar{\xi}}(4,1) = \sum_{i=1}^{4} \xi_i e^{\theta(t,x,y;k_i)} \quad . \tag{24}$$

The solution contains four external solitons. Depending on the wave numbers, it describes the splitting of one soliton into three, or a collision involving two incoming and two outgoing solitons. A splitting process is presented at different times in Figs. 3-5 for $\xi_i = 1$, $1 \leq i \leq 4$. At $t = 0$ (Fig. 3), there is one junction. Two junctions, connected by a single soliton, emerge away from $t = 0$.

The asymptotic structure of the $(M,N) = (M,1)$ solutions with $M \geq 4$ (equivalently, the space-time-inversion related $(M, N-1)$ solutions) has been elucidated in the literature through the dominant-phase approach [31-44]. The usual methods for wave-number allocation to solitons are described in Appendix I. In order to focus on vertex dynamics, the case of $M = 4$ is analyzed from a slightly different viewpoint - of "riding" on each Y-shaped junction and exploiting the fact that, as $|t| \to \infty$, each junction propagates at a constant velocity, given by Eq. (20).

With four wave numbers, one can form four triplets, hence, potentially, four Y-shaped segments may exist. To see which of them persists as $|t| \to \infty$, one computes $u(t, x+v_x t, y+v_y t)$ with the velocity (Eq. (20)) computed for each triplet of wave numbers. For $0 < k_1 < k_2 < k_3 < k_4$, Eqs. (24), (5) and (4) yield the following limits:

$$u(t, x + v_x(k_1,k_2,k_3)t, y + v_y(k_1,k_2,k_3)) \xrightarrow[t \to +\infty]{} Y(x,y;1,2,3)$$
$$u(t, x + v_x(k_1,k_3,k_4)t, y + v_y(k_1,k_3,k_4)) \xrightarrow[t \to +\infty]{} Y(x,y;1,3,4)$$
$$u(t, x + v_x(k_1,k_2,k_3)t, y + v_y(k_1,k_2,k_3)) \xrightarrow[t \to -\infty]{} 0$$
$$u(t, x + v_x(k_1,k_3,k_4)t, y + v_y(k_1,k_4,k_4)) \xrightarrow[t \to -\infty]{} 0$$

(25)

$$u(t, x + v_x(k_2,k_3,k_4)t, y + v_y(k_2,k_3,k_4)) \xrightarrow[t \to -\infty]{} Y(x,y;2,3,4)$$
$$u(t, x + v_x(k_1,k_2,k_4)t, y + v_y(k_1,k_2,k_4)) \xrightarrow[t \to -\infty]{} Y(x,y;1,2,4)$$
$$u(t, x + v_x(k_2,k_3,k_4)t, y + v_y(k_2,k_3,k_4)) \xrightarrow[t \to +\infty]{} 0$$
$$u(t, x + v_x(k_1,k_2,k_4)t, y + v_y(k_1,k_2,k_4)) \xrightarrow[t \to +\infty]{} 0$$

(26)

In Eqs. (25) and (26), $Y(x,y;i,j,l)$ denotes a Y-shaped solution of the KP II equation, with wave numbers $k_i$, $k_j$ and $k_l$, and coefficients, $\xi_i$, $\xi_j$ and $\xi_l$, evaluated at $t = 0$.

As $|t| \to \infty$, the (4,1) solution evolves into two Y-shaped segments, each around one junction. Hence, each segment moves at the constant velocity of Eq. (20), computed with the appropriate wave numbers. The distance between the two junctions grows indefinitely for $|t| \gg 0$. (From Eq. (20), one finds that the two junctions do not move apart only when the solution degenerates into a single soliton with $k_4 = k_2$ and $k_3 = k_1$.) At large $|t|$, the two segments are connected by a single soliton, which is constructed from the two wave numbers that are common to both segments. The segments at $t \to +\infty$ and $t \to -\infty$ are not identical. As a result, the connecting soliton is constructed from different wave-number pairs in the two limits: $\{k_1, k_3\}$ at $t \gg 0$, and $\{k_2, k_4\}$ at $t \ll 0$.

The change in wave-number allocation to single solitons is a manifestation of *change in the structure of the solution under space-time inversion* (see Section 2.1.2). The other manifestation is the fact that its $t \gg 0$ and $t \ll 0$ graphs are not mirror images of one another. Direct inspection of the solution reveals that all solitons meet within one junction at a time, which depends on the values of $\xi_i$. For example, when all $\xi_i = 1$, this happens around $t = 0$. An example is shown in Figs. 3-5.

### 2.2.3.2 Four wave numbers – four soliton junctions

The (4,2) solution is generated by:

$$f(t,x,y) = \sum_{1 \leq i < j \leq 4} \xi_{ij}(k_j - k_i) e^{\theta(t,x,y;k_i) + \theta(t,x,y;k_j)} \qquad (k_1 < k_2 < k_3 < k_4) \ . \tag{27}$$

The coefficients $\xi_{ij}$ must be positive and are constrained by Eq. (9).

Except for specific choices of the $\xi$-coefficients, the solution generated from Eq. (27) contains four junctions. An analysis of the solution along the lines of Section 2.2.3.1 reproduces the results of [33], which are summarized in Figs. 6-8. As $|t| \to \infty$, the solution exhibits four soliton junctions (Figs. 6 and 8). Near $t = 0$, the four junctions coalesce into one junction (Fig. 7). Based on the discussion in Section 2.1.2, apart from constant finite shifts, the graphs of the solution at $t \to \pm \infty$ are mirror images of one another, since their structure is invariant under space-time inversion.

### 3. Vertex maps
### 3.1 Single-soliton identity

For a single-soliton solution, $u(t,x,y) = f(\xi = x + a y + \omega t)$, Eq. (3) becomes:

$$(3a^2 - 4\omega) f'' + 6(f'^2 + f f'') + f'''' = 0 \ . \tag{28}$$

Employing the vanishing boundary condition at $\xi = -\infty$, two identities are obtained:

$$(3a^2 - 4\omega) f + 3f^2 + f'' = 0 \ , \qquad \frac{1}{2}(3a^2 - 4\omega) f^2 + 2f^3 - \frac{1}{2} f'^2 + f f'' = 0 \ . \tag{29}$$

Eqs. (29) yield an identity that is independent of $a$ and $\omega$:

$$f^3 + f f'' - (f')^2 = 0 \ . \tag{30}$$

Hence, the vanishing of the differential polynomial,

$$R[u] = u^3 + u u_{xx} - (u_x)^2 \ , \tag{31}$$

when $u$ is a single-KP II-soliton solution, is a direct consequence of the KP II equation, Eq. (3).

On multi-soliton solutions, $R[u]$ vanishes along single-soliton lines far from the soliton junctions, and, hence, generates a map of localized structures, one for each junction. This statement will be now demonstrated in the cases of the solutions discussed in Section 2.

**3.2 Free single vertices**
**3.2.1 Three wave numbers: *Y*-shaped solution**
Substituting Eq. (18) for $f(t,x,y)$ in Eq. (4), and the resulting expression for the solution, $u(t,x,y)$, in Eq. (31), one finds that the image of the *Y*-shaped solution under $R[u]$ is:

$$R[u_Y] = \frac{4 e^{\theta_1(t,x,y) + \theta_2(t,x,y) + \theta_3(t,x,y)}}{\left(e^{\theta_1(t,x,y)}\xi_1 + e^{\theta_2(t,x,y)}\xi_2 + e^{\theta_3(t,x,y)}\xi_3\right)^3} (k_1 - k_2)^2 (k_1 - k_3)^2 (k_2 - k_3)^2 \xi_1 \xi_2 \xi_3 \, , \qquad (32)$$

where $\theta_i(t,x,y)$ are defined in Eq. (7). The subscript *Y* has been appended to denote the fact that this is the image of the *Y*-shaped solution.

$R[u_Y]$ of Eq. (32) has a maximal value at a point within the junction, and decays exponentially in all directions away from the soliton junction. An example is shown in Fig. 9. The maximum is:

$$\max[R[u_Y(t,x,y)]] = \frac{4}{27}(k_1 - k_2)^2 (k_1 - k_3)^2 (k_3 - k_2)^2 \ \left(x = v_x t + \delta_x, y = v_y t + \delta_y\right) \, . \qquad (33)$$

(In Eq. (33), $v_x$ and $v_y$ are given by Eq. (20), and $\delta_x$ and $\delta_y$ depend on $k_i$ and $\xi_i$, $1 \leq i \leq 3$. For example, when all $\xi_i$ are equal, $\delta_x = \delta_y = 0$.)

To expose the exponential decay, consider one of the single solitons that emerge from the junction. Denote the wave numbers, from which it is constructed, by $k_i$ and $k_j$ ($1 \leq i \neq j \leq 3$) and the third wave number (which participates in the construction of the other two solitons) by $k_l$, $l \neq i,j$. Now, introduce orthogonal coordinates, $\xi$ and $\eta$, where $\xi$ measures the distance from the junction along the soliton trajectory, and $\eta$ measures the distance from the soliton trajectory in a direction perpendicular to the trajectory (it is the argument in the solution of a single soliton, see Eq. (17)):

$$\xi = \left(k_i + k_j\right)x + y \quad , \qquad \eta = x - \left(k_i + k_j\right)y + \left(k_i^2 + k_i k_j + k_j^2\right)t \quad . \tag{34}$$

From Eq. (32), one finds that at any fixed time, $t$, for large $|\xi|$ and $|\eta|$, $R[u_Y]$ falls off as:

$$R[u_Y] \propto e^{-\alpha|\xi|} e^{-\beta|\eta|} \quad . \tag{35}$$

In Eq. (35), one has:

$$\alpha = \left|\left(k_l - k_i\right)\left(k_l - k_j\right)\right| \Big/ \left(1 + \left(k_i + k_j\right)^2\right)$$

$$\beta = \max\left\{\left|\left(k_l - k_i\right)\left(1 + \left(k_i + k_l\right)\left(k_i + k_j\right)\right)\right|, \left|\left(k_l - k_j\right)\left(1 + \left(k_j + k_l\right)\left(k_i + k_j\right)\right)\right|\right\} \Big/ \left(1 + \left(k_i + k_j\right)^2\right) \tag{36}$$

Thanks to Eq. (19), the single vertex, generated by $R[u_Y]$ of Eq. (32), preserves its shape in time as it rigidly propagates at the constant velocity given by Eqs. (20). Hence, one may view it as emulating a free, spatially extended, particle. To complete the particle analogy, a vertex has to be assigned a mass. The definition, commonly employed in soliton dynamics, of an integral of the structure over space, is adopted. Direct integration yields in the case of $k_3 > k_2 > k_1$:

$$m_{VERTEX} = \int_{-\infty}^{\infty}\int_{-\infty}^{\infty} R\left[u_Y(t,x,y)\right]dx\,dy = 2\left(k_2 - k_1\right)\left(k_3 - k_1\right)\left(k_3 - k_2\right) \quad . \tag{37}$$

The mass, Eq. (37), vanishes if any two wave numbers are equal. This is a reflection of the fact that $R[u_Y]$ exhibits the same behavior: The $Y$-shaped solution then degenerates into a single soliton, for which $R[u]$ vanishes.

A valid question at this point is whether there is any justification, besides intuition, for the definition of the mass as the spatial integral, Eq. (37). An analysis, presented in Appendix II, shows that, up to a multiplicative constant, Eq. (37) is the only definition of vertex mass, which ensures momentum conservation in the collision processes generated from $(M,1)$ solutions for all $M \geq 4$.

### 3.2.2 *X*-shaped solution

The *X*-shaped solution preserves its spatial structure in time, as it propagates at a constant velocity [15, 31-41]. (For the $\xi$-coefficients of Eq. (21), the velocity is given by Eq. (22)). Consequently, its single vertex map, generated by $R[u]$ of Eq. (31), emulates a free, spatially extended, particle. An example is shown in Fig. 10. Repetition of the analysis along the lines of Eqs. (33)-(36), leads to identical results. $R[u_X]$ falls off exponentially fast away from the junction, and has a non-vanishing maximum at a point within the junction. Adopting the definition as in Eq. (37), the "particle" mass is found to be:

$$m_{VERTEX} = \int_{-\infty}^{\infty}\int_{-\infty}^{\infty} R\left[u_X(t=0,x,y)\right]dx\,dy = \frac{4|k_1 - k_2||k_3 - k_4|(k_1 - k_2 + k_3 - k_4)^2}{|(k_1 + k_2) - (k_3 + k_4)|} \quad . \tag{38}$$

The mass becomes infinite when the denominator in Eq. (38) vanishes. However, as noted in Section 2.2.2.2, in this limit, the *X*-type solution degenerates into two parallel solitons. Furthermore, in solutions constructed with $M > 4$ wave numbers, *X*-shaped segments do not persist as $|t| \to \infty$ [37], hence, are not discussed in the remainder of the paper.

### 3.3 Collision processes
Most $M \geq 4$ solutions of Eq. (3) contain several soliton junctions. The junctions move in time, and are well separated at $|t| \gg 0$. Their images under $R[u]$ of Eq. (31) describe collision processes amongst several vertices. The number of vertices equals the number of soliton junctions displayed by the solution at $|t| \gg 0$. Whether vertex collisions are elastic or inelastic depends solely on the large-time properties of the soliton solutions under space-time inversion. The collisions are elastic in the case of $M = 2\,N$ solutions and inelastic for solutions with $M \neq 2\,N$.

### 3.3.1 Elastic collisions
Up to constant shifts, the structure of solutions with $M = 2\,N$ at $\{t,x,y\}$ is identical to their structure at $\{-t, -x, -y\}$. Specifically, wave-number assignment to solitons is the same in both cases (see Section 2.1.2). Figs. 6 and 8 of an $(M,N) = (4,2)$ solution of Eq. (27) provide an example. The temporal evolution of the four vertices generated from the $(M,N) = (4,2)$ solution, shown in Figs. 11-13, follows this pattern. At large $|t|$, the map shows four vertices. As $t \to 0$, the four collide and coalesce into one vertex. For $|t| \gg 0$, when the distances amongst junctions are much greater than the finite shifts, vertex velocities and masses in the vertex map do not change under space-time inversion. As a result, these vertex maps, trivially, emulate elastic collisions.

### 3.3.2 Inelastic collisions

For $M \neq 2N$, the structure of a solution at $\{t,x,y\}$ is different from its structure at $\{-t, -x, -y\}$. Specifically, wave-number assignment to solitons is changed (see Section 2.1.2). Consequently, in general, the vertex maps of these solutions emulate inelastic collisions. The masses (e.g., Eq. (37)) and velocities (e.g. Eq. (20)) of vertices at $t \gg 0$ and at $t \ll 0$ are different. For wave numbers $\{k_1, k_2, k_3, k_4\} = \{1, \sqrt{3}, \sqrt{7}, \sqrt{11}\,\}$, vertex masses and velocities change as follows:

$$
\begin{aligned}
&m_I^{in} = 5.3745 \quad m_{II}^{in} = 1.9426 \quad m_I^{out} = 5.1155 \quad m_{II}^{out} = 2.2016 \\
&v_{I,x}^{in} = 10.7932 \quad v_{II,x}^{in} = 19.1021 \quad v_{I,x}^{out} = 24.7373 \quad v_{II,x}^{out} = 8.9604 \\
&v_{I,y}^{in} = 6.0487 \quad v_{II,y}^{in} = 7.6944 \quad v_{I,y}^{out} = 6.9624 \quad v_{II,y}^{out} = 5.3778
\end{aligned} \quad (39)
$$

Here, $I$ and $II$ correspond to the notation in Figs. 4 and 5, and $\{in, out\}$ correspond to $\{t \ll 0, t \gg 0\}$.

The vertex maps of the multi-soliton solution with $(M,N) = (4,1)$ of Figs. 3-5 are shown in Figs. 14-16. The vertices undergo an inelastic collision. At $|t| \gg 0$, the vertex maps follow the asymptotic structure of the solution. They exhibit two vertices that move in the $x$-$y$ plane. At some finite time ($t = 0$ in the numerical example, in which all $\xi_i = 1$) the two vertices coalesce (see Fig. 15).

## 4. Collision kinematics

Upon collision, the vertices lose their identities and coalesce into one vertex. Therefore, the kinematic analysis is restricted, as commonly done in the phenomenological study of particle collisions, to the behavior of the system as $|t| \to \infty$, when vertices are far apart. Whether the collisions are elastic or inelastic depends *solely* on the properties of the asymptotic structure of the soliton web under space-time inversion, $(t,x,y) \to (-t, -x, -y)$. Vertex velocities are determined by Eq. (20). However, the definition of vertex-mass remains an open question. Eq. (37) is still an intuitively motivated choice.

## 4.1 Elastic collisions – No constraints on vertex mass

Vertices generated from solutions with $(M,N) = (2N,N)$ undergo elastic collisions: vertex masses and velocities are not changed through the collision. The velocities are determined by the soliton solution (see Eq. (20)). However, there are no constraints on vertex masses.

## 4.2 Inelastic collisions - Momentum conservation determines vertex mass

Vertices generated from solutions with $(M,N)$ with $M \neq 2N$, undergo inelastic collisions. Within this subset of solutions, the infinite family of $(M,1)$ solutions, with $M \geq 4$ offers a way to determine vertex masses that allow for the "particle" interpretation of the colliding vertices in the case of *Y*-shaped junctions.

The Young-diagram- and permutations-analysis of [17-22, 31-37] and the Tamari-lattice analysis of [42-44], yield that, at $|t| \to \infty$, the soliton web generated by an $(M,1)$ solution with an ordered set of wave numbers (Eq. (6)) resembles a comb-like structure, containing $(M–2)$ *Y*-shaped junctions. The junctions are constructed from the following triplets of wave numbers:

$$\begin{aligned} \text{In:} \quad & \left(k_{M-2}, k_{M-1}, k_M\right), \left(k_{M-3}, k_{M-2}, k_M\right), \ldots, \left(k_1, k_2, k_M\right) & t \to -\infty \\ \text{Out:} \quad & \left(k_1, k_{M-1}, k_M\right), \left(k_1, k_{M-2}, k_{M-1}\right), \ldots, \left(k_1, k_2, k_3\right) & t \to +\infty \end{aligned} \quad (40)$$

As the distances amongst the junctions are large, each is mapped by Eq. (31) onto a single vertex. The (4,1) solution, presented in Figs. 3–5, is an example to this rule.

In solutions with $M \geq 5$, the number of unknown masses, equal to the number of junctions, is $2 \cdot (M–2) \geq 6$. The small number of conservation laws is insufficient for the determination of the wave-number dependence of this number of masses. The problem is resolved by imposing momentum conservation on the vertex-collision process generated from the (4,1) solution, because there are then only four unknown masses. The detailed analysis is carried out in Appendix II. It exploits the fact that, as $|t| \to \infty$, the solution evolves into two distinct *Y*-shaped junctions, which do not affect one another. The wave-number dependence of a vertex mass is required to be such that: 1) It is non-singular; 2) it is the same for all wave numbers; and 3) the mass vanishes when any two wave numbers coincide. The conclusions are presented in the following.

Linear momentum conservation

The only definition of the mass, consistent with momentum conservation, is:

$$m(k_1, k_2, k_3) = \alpha (k_3 - k_1)(k_3 - k_2)(k_2 - k_1) \quad . \tag{41}$$

The significance of Eq. (41) is that factorization of the solution into *Y*-shaped segments, which do not affect one another, combined with momentum conservation, are only consistent with a vertex mass that, up to a multiplicative constant, must be the spatial integral of Eq. (37), i.e., to interpreting the profile of a structure as the *mass density of a spatially extended particle*.

Conserved quantities, $M \geq 4$

With Eq. (41) for vertex mass, and using Eq. (20) for vertex velocity, simple algebra yields the following conservation laws for the infinite family of (*M*,1) solutions for any $M \geq 4$:

Total mass:

$$M_{in} = \sum_{i=1}^{M-2} m(k_i, k_{i+1}, k_M) = M_{out} = \sum_{i=1}^{M-2} m(k_1, k_{i+1}, k_{i+2}) \quad , \tag{42}$$

Linear momentum:

$$\begin{aligned} P_{x,in} &= \sum_{i=1}^{M-2} m(k_i, k_{i+1}, k_M) v_x(k_i, k_{i+1}, k_M) = P_{x,out} = \sum_{i=1}^{M-2} m(k_1, k_{i+1}, k_{i+2}) v_x(k_1, k_{i+1}, k_{i+2}) \\ P_{y,in} &= \sum_{i=1}^{M-2} m(k_i, k_{i+1}, k_M) v_y(k_i, k_{i+1}, k_M) = P_{y,out} = \sum_{i=1}^{M-2} m(k_1, k_{i+1}, k_{i+2}) v_y(k_1, k_{i+1}, k_{i+2}) \end{aligned}, \tag{43}$$

The conservation of the total mass, Eq. (42), and of the total linear momentum, Eqs. (43), ensures that the velocity of the center of mass is also unchanged through the collision process.

Finally, the kinetic energy due to motion along the *y*-direction is also conserved:

$$K_{y,in} = \frac{1}{2} \sum_{i=1}^{M-2} m(k_i, k_{i+1}, k_M) \left( v_y(k_i, k_{i+1}, k_M) \right)^2 = K_{y,out} = \frac{1}{2} \sum_{i=1}^{M-2} m(k_1, k_{i+1}, k_{i+2}) \left( v_y(k_1, k_{i+1}, k_{i+2}) \right)^2 \quad .\tag{44}$$

However, in general, the total kinetic energy due to the motion along the $x$ – axis is not conserved:

$$K_{x,in} = \frac{1}{2}\sum_{i=1}^{M-2} m(k_i, k_{i+1}, k_M)(v_x(k_i, k_{i+1}, k_M))^2 \neq K_{x,out} = \frac{1}{2}\sum_{i=1}^{M-2} m(k_1, k_{i+1}, k_{i+2})(v_x(k_1, k_{i+1}, k_{i+2}))^2 \quad .(45)$$

Total kinetic energy conservation is possible either when all wave numbers coincide (the solution then vanishes), or if they obey some relation. In the case of the (4,1) solution, that relation is:

$$k_1 + k_2 + k_3 + k_4 = 0 \quad . \tag{46}$$

Exploiting Eqs. (20), one finds that, for a genuine (4,1)-solution, vertex velocities are changed in the collision. Forcing the velocities to remain the same causes the solution to degenerate into a single-soliton solution. The masses (Eq. (37)) are unchanged in this trivial case, but also in a nontrivial one, when the wave numbers obey one of the following conditions:

$$\begin{aligned} k_1 + k_4 &= k_2 + k_3 \\ k_1 + k_2 &= k_3 + k_4 \end{aligned} \qquad (k_4 > k_3 > k_2 > k_1) \quad . \tag{47}$$

## 5. Unique role of $R[u]$

There is an infinite hierarchy of differential polynomials, which, like $R[u]$ of Eq. (31), vanish on the single-soliton solution, and map multi-soliton solutions onto a system of vertices. The existence of such a hierarchy has been discussed in detail in the case of the KdV equation [47]. These "special polynomials" can be classified according to scaling weight, $W$. Assigning $\{u, \partial_t, \partial_x, \partial_y\}$ the scaling weights $\{2, 3, 1, 2\}$, every term in the KP II equation, Eq. (3), and in $R[u]$ has $W = 6$.

However, $R[u]$ plays a unique role. To begin with, an infinite number of these special polynomials are non-local. Namely, they contain integrals of monomials involving $u$ and/or its spatial derivatives. In fact, the first special polynomial has $W = 3$ and is non-local. It can be obtained from the vanishing of $R[u]$ (Eq. (31)) on a single-soliton solution:

$$u + \left((u u_{xx} - (u_x)^2)/u^2\right) = u + \partial_x(u_x/u) = 0 \quad . \tag{48}$$

Integration over $x$ yields that the $W=3$ non-local differential polynomial,

$$R^{(3)} = u \partial_x^{-1} u + u_x \quad , \tag{49}$$

vanishes on the single-soliton solution, provided the integrator is defined as:

$$\partial_x^{-1} u = \frac{1}{2}\left( \int_{-\infty}^{x} u\, dx - \int_{x}^{+\infty} u\, dx \right) . \tag{50}$$

Owing to its non-local nature, $R^{(3)}$ does not map a multi-soliton solution into a collection of vertices, i.e., distinct spatially localized structures. For example, in the case of the Y-shaped solution of Eq. (18), with $0 < k_1 < k_2 < k_3$ (see Fig. 1), a detailed calculation yields that, at any time, away from the junction, $R^{(3)}$ tends to:

$$R^{(3)} \to \begin{cases} 0 & (x \ll v_x t, y \ll v_y t) \\ (k_2 - k_1) u_{Single}(k_2, k_3) - (k_3 - k_2) u_{Single}(k_1, k_2) & (x \gg v_x t, y \gg v_y t) \end{cases} . \tag{51}$$

In Eq. (51), $u_{Single}(k_1, k_2)$ is the single soliton generated with wave numbers $(k_1, k_2)$, $v_x$ and $v_y$ are the velocity components given in Eqs. (20), and the limitations on $x$ and $y$ ensure that one is far away from the soliton junction. Thus, $R^{(3)}$ generates long-range interactions among solitons. At higher values of $W$, the number of non-local special polynomials grows rapidly.

The first local special polynomial is $R[u]$ of Eq. (31) ($W=6$). At higher values of $W$, the number of local special polynomials also grows rapidly. As an example for the construction of these polynomials, the case $W=8$.

One writes down the most general differential polynomial with $W = 8$. For it to be local, it may contain only powers of $u$ and of its spatial derivatives:

$$\begin{aligned} R^{(8)} = &\, a_1 u^4 + a_2 u^2 u_{xx} + a_3 u (u_x)^2 + a_4 u^2 u_y + a_5 u u_{xxxx} + a_6 u_x u_{xxx} \\ &+ a_7 (u_{xx})^2 + a_8 u u_{xxy} + a_9 u_x u_{xy} + a_{10} u_{xx} u_y + a_{11} u u_{yy} + a_{12} (u_y)^2 \end{aligned} . \tag{52}$$

One now requires that $R^{(8)}$ vanish on the single-soliton solution of Eq. (17). This turns out to leave only five coefficients out of $a_{1-12}$ independent, so that there are five linearly independent local special polynomials with scaling weight $W = 8$:

$$\begin{aligned}
R^{(8,1)}[u] &= u\,R[u] \\
R^{(8,2)}[u] &= \partial_x^{\,2} R[u] \\
R^{(8,3)}[u] &= u^2 u_{xx} - 3u(u_x)^2 + (u_{xx})^2 - u_x u_{xx} \\
R^{(8,4)}[u] &= \partial_y R[u] \\
R^{(8,5)}[u] &= u_{xx} u_y - u_x u_{xy}
\end{aligned} \qquad (53)$$

The unique role played by $R[u]$ of Eq. (31) now emerges. Local polynomials with $W > 6$ do generate their own vertices, i.e., structures that are localized around soliton junctions, and move at velocities given, by Eqs. (20). However, to begin with, amongst the five $W = 8$ polynomials, only $R^{(8,1)}[u]$ is positive definite. Hence, it is the only candidate for a mass density. Furthermore, if the mass of a vertex generated from a $W > 6$ polynomial is defined through the spatial integral of in Eq. (37), then its wave number-dependence is different from the cubic result of Eq. (37). For example, replacing $R[u]$ in Eq. (37) by any of the five polynomials with $W = 8$ in the case of the Y-shaped solution of Eq. (18) generates a quintic dependence on the wave numbers. With such masses, none of the conservation laws of Eqs. (42)-(44) are obeyed.

**6. Vertex dynamics**
The vertices are not point-like, and their properties change as the distances amongst them approach their own spatial extent. They then lose their identities, and, eventually, coalesce into one vertex. Hence, a description in terms of point-like particles that move in a potential cannot be applied. The following is a report of preliminary results obtained in the search for a theory of vertex dynamics. Exploiting Eq. (3), the temporal evolution of $R[u]$ of Eq. (31) is found to be:

$$\partial_t R[u] = \partial_x \left\{ \frac{3}{2} R^{(8,1)}[u] + \tilde{R}[u] \right\} , \qquad (54)$$

$$\tilde{R}[u] = \frac{3}{4}\partial_x^{-1}\left\{\left(\partial_x^{-1}\left(\partial_y^2 u\right)\right)\left\{3u^2 + u_{xx}\right\}\right\} - \frac{9}{4}\partial_x^{-1}\left(u_x u_{yy}\right) + \frac{3}{4}u u_{yy}$$
$$+ \frac{9}{4}u^4 + \frac{15}{4}u^2 u_{xx} - 3u\left(u_x\right)^2 + \frac{1}{2}\left(u_{xx}\right)^2 - \frac{3}{4}u_x u_{xxx} + \frac{1}{4}u u_{xxxx}$$
(55)

In Eq. (55), the integrator is defined as in Eq. (50).

$\tilde{R}[u]$ is a scaling-weight $W = 8$ differential polynomial, which vanishes on a single-soliton solution, but is not expressible in terms of $R[u]$. It contains non-local terms, in which solitons have long-range effects on one another. In addition, the difficulties discussed in Section 5 arise. Its space integral cannot serve as a mass that obeys the conservation laws of Eqs. (42)-(44). Hence, Eq. (54) couples one family of vertices to another family. As there is an infinite hierarchy of differential polynomials that vanish on a single-soliton solution, the family of vertices generated by one special polynomial is coupled to families of vertices generated by other special polynomials.

## 7. Concluding comments

In this paper, the dynamics of vertices in vertex maps of multi-soliton solutions of the KP II equation in (1+2) dimensions has been studied. A functional of the solution, which vanishes on single solitons, generates vertex maps of multi-soliton solutions. The vertices are images of soliton junctions. When far apart, they emulate free, spatially extended, non-relativistic particles.

In solutions whose structure changes under space-time inversion, vertex collisions are inelastic. In particular, in the $(M,N) = (M,1)$ solutions (see Section 4.2.1), the kinetic energy due the motion in the $y$ direction is always conserved, but the kinetic energy due the motion in the $x$ direction is not conserved in general. While there is yet no rigorous explanation for this asymmetry it must, obviously, be related to the anisotropic nature of the KP II equation, Eq. (3). Conservation in the $y$ direction is, most probably, a consequence of the Laplacian form of the $y$-term in Eq. (3) ($\partial_y^2 u$).

The dissipative nature of the part of Eq. (3) that contains derivatives with respect to *x*, is, most probably, the cause for the lack of conservation in the *x* direction.

In soliton dynamics, the mass of a localized structure has been often defined as its spatial integral (see, e.g., Eqs. (37) and (38)). The analysis presented here provides an example of a system, for which this definition is essential for the interpretation of the vertices as colliding spatially extended non-relativistic particles; it is intimately related with the requirement of momentum conservation. Finally, similar hidden "particle" dynamics is expected to be uncovered in cases of other integrable nonlinear evolution equations in more than one space dimension that have line-soliton solutions. A local functional of the solution, which vanishes on a single-soliton, will generate localized structures in the vicinity of soliton junctions.

Acknowledgment The author is greatly indebted to F. Müller-Hoissen for careful reading of the manuscript, as we well as for thorough, incisive and constructive criticism.

**Appendix I. Wave numbers-assignment to solitons in multi-soliton solutions of KP equation**

In the following, three tools that assist in the assignment of wave numbers to solitons are described. All three tools have been employed in the analysis of the solutions discussed in the paper.

First, a rigorous way to assign wave-number pairs to solitons has been elucidated in [31-43], exploiting the fact that the interplay between the phases, $\theta(t,x,y,k_i)$ ($1 \leq i \leq 3$) (see Eq. (5)), determines wave-number assignment. For each phase, $\theta(t,x,y,k_i)$, one identifies the "dominance" domain $D$ of this phase in the $x$-$y$ plane, i.e., the domain, in which it exceeds all other phases:

$$\theta(t,x,y,k_i) > \theta(t,x,y,k_j) \quad (x,y) \in D, \quad 1 \leq j \neq i \leq 3 \ . \tag{I.1}$$

Soliton lines are located along the straight-line boundaries that separate dominance regions of different phases. For example, the line, which tends to a soliton constructed from wave numbers $k_1$ and $k_2$, is the boundary between the dominance regions of $\theta(t,x,y,k_1)$ and $\theta(t,x,y,k_2)$.

Next, at the maximum of the profile of the single-soliton solution, Eq. (17), (wave numbers $k_i$ and $k_j$), one has:

$$\partial_x^2 u(t,x,y)\Big|_{\text{At maximum}} = -(k_i - k_j)^4/4 \ , \quad \partial_y^2 u(t,x,y)\Big|_{\text{At maximum}} = -\left((k_i - k_j)^4 (k_i + k_j)^2\right)/4 \ . \tag{I.2}$$

In the $x$–$y$ plot of a multi-soliton solution at some time $t$, one focuses on one of the solitons away from the collision vertex. Using Eq. (I.2), the numerical values of the derivatives in the actual solution uniquely determine the wave numbers associated with the soliton.

Finally, a consistency check of the wave-number assignment exploits the fact that, at the point of maximum along the trajectory of a soliton that is constructed from wave numbers $k_i$ and $k_j$, the difference $\{\theta(t,x,y,k_i) - \theta(t,x,y,k_j)\}$ vanishes. One or more of the differences, $\{\theta(t,x,y,k_m) - \theta(t,x,y,k_l)\}$, for all $l \neq m$, vanish at the point of maximum, the pair $\{m = i, l = j\}$ included.

**Appendix II. Momentum conservation and definition of vertex mass - (4,1) solution**

The vertex map of a (4,1) solution contains four unknown masses (see Figs.14 and 16). The incoming and outgoing masses are different. Imposing momentum conservation, Eqs. (43), on the vertex map, and using Eq. (20) for vertex velocity, one can express two of the masses in terms of the other two:

$$m[k_1,k_2,k_3] = \frac{(k_3-k_2)}{(k_4-k_3)} \frac{(k_1^2+k_1 k_4+k_4^2)}{(k_1^2+k_1 k_2+k_2^2)} m[k_1,k_3,k_4]$$

$$- \frac{(k_3-k_1)}{(k_4-k_3)} \frac{(k_2^2+k_2 k_4+k_4^2)}{(k_1^2+k_1 k_2+k_2^2)} m[k_2,k_3,k_4]$$

$$m[k_1,k_2,k_4] = \frac{(k_4-k_2)}{(k_4-k_3)} \frac{(k_1^2+k_1 k_3+k_3^2)}{(k_1^2+k_1 k_2+k_2^2)} m[k_1,k_3,k_4]$$

$$- \frac{(k_3-k_1)}{(k_4-k_3)} \frac{(k_2^2+k_2 k_3+k_3^2)}{(k_1^2+k_1 k_2+k_2^2)} m[k_2,k_3,k_4]$$

. (II.1)

$m[k_1,k_2,k_3]$ cannot depend on $k_4$, and $m[k_1,k_2,k_4]$ cannot depend on $k_3$. This yields two constraints:

$$\partial_{k_4} m[k_1,k_2,k_3] = (k_4-k_3)^2 (k_3-k_1)(k_3-k_2) \times$$
$$\left\{ \frac{\partial_{k_4}\left(\frac{(k_4^3-k_1^3)}{(k_4-k_1)(k_4-k_3)} m[k_1,k_3,k_4]\right)}{(k_3-k_1)} - \frac{\partial_{k_4}\left(\frac{(k_4^3-k_2^3)}{(k_4-k_2)(k_4-k_3)} m[k_2,k_3,k_4]\right)}{(k_3-k_2)} \right\} = 0$$

, (II.2)

$$\partial_{k_3} m[k_1,k_2,k_4] = (k_4-k_3)^2 (k_4-k_1)(k_4-k_2) \times$$
$$\left\{ \frac{\partial_{k_3}\left(\frac{(k_3^3-k_1^3)}{(k_4-k_3)(k_3-k_1)} m[k_1,k_3,k_4]\right)}{(k_4-k_1)} - \frac{\partial_{k_3}\left(\frac{(k_3^3-k_2^3)}{(k_4-k_3)(k_3-k_2)} m[k_2,k_3,k_4]\right)}{(k_4-k_2)} \right\} = 0$$

. (II.3)

Let us focus on Eq. (II.2). The first and second terms inside the curly brackets must be independent of $k_1$, and $k_2$, respectively:

$$\partial_{k_1}\left\{\left[\partial_{k_4}\left(\frac{(k_3^3 - k_1^3)}{(k_4 - k_3)(k_3 - k_1)} m[k_1, k_3, k_4]\right)\right]\middle/(k_4 - k_1)\right\} =$$
$$\partial_{k_2}\left\{\left[\partial_{k_4}\left(\frac{(k_3^3 - k_2^3)}{(k_4 - k_3)(k_3 - k_2)} m[k_2, k_3, k_4]\right)\right]\middle/(k_4 - k_2)\right\} = 0$$

(II.4)

The two constraints of Eq. (II.4) yield two different solutions for the dependence of the mass on the three wave numbers. These must coincide, hence, denoted temporarily by $m_1$ and $m_2$:

$$m_1[p,r,s] = \frac{(s-r)(r-p)(s-p)}{s^3 - p^3}\int_p^s C_1[r,x]dx + \frac{(s-r)(s-p)}{s^3 - p^3}C_2[p,r]$$
$$m_2[p,r,s] = \frac{(s-r)(r-p)(s-p)}{r^3 - p^3}\int_p^r C_3[s,x]dx + \frac{(s-r)(r-p)}{r^3 - p^3}C_4[p,s]$$

(II.5)

In Eqs. (II.5), $C_{1-4}$ are arbitrary functions. Requiring that both expressions vanish when any two wave numbers coincide, yields:

$$C_2[p,r] = C_4[p,s] = 0 \ .$$ (II.6)

Next, one requires that the two different factors in the remainder of Eqs. (II.5) coincide:

$$\left(\int_p^s C_1[r,x]dx\right)\middle/(s^3 - p^3) = \left(\int_p^r C_3[s,x]dx\right)\middle/(r^3 - p^3) \ .$$ (II.7)

Requiring Eq. (II.7) to hold, in particular, in the limit $s \to r$, yields

$$\int_p^r C_1[r,x]dx = \int_p^r C_3[r,x]dx \ ,$$ (II.8)

which implies

$$C_1[r,x] = C_3[r,x] \ .$$ (II.9)

Eq. (II.7) can be now re-written as:

$$\left(r^3 - p^3\right)\int_p^s C_1[r,x]dx = \left(s^3 - p^3\right)\int_p^r C_1[s,x]dx \quad . \tag{II.10}$$

The only solution to this constraint is:

$$C_1[y,x] = \alpha\, 3x^2 \quad , \tag{II.11}$$

with $\alpha$ - a constant. Eqs. (II.5)-(II.11) now yield a unique expression for the mass:

$$m[p,r,s] = \alpha(s-r)(r-p)(s-p) \quad . \tag{II.12}$$

Using Eq. (II.12), one finds that the constraint of Eq. (II.3), which has not been addressed hitherto, is obeyed automatically.

Eq. (II.12) leads to the important conclusion that, up to a multiplicative constant, momentum conservation is possible only with a vertex mass that coincides with the spatial integral of Eq. (37).

Finally, the mass definition of Eq. (II.12) yields that all three conservation laws, Eqs. (42)-(44), are automatically obeyed for all $M \geq 4$. The proof is simple. As an example, it is given here for the conservation of the total mass.

$$\begin{aligned} M_{in} - M_{out} &= \sum_{i=1}^{M-2} m\left[k_i, k_{i+1}, k_M\right] - \sum_{i=1}^{M-2} m\left[k_1, k_{i+1}, k_{i+2}\right] \\ &= m\left[k_1, k_2, k_M\right] - m\left[k_1, k_{M-1}, k_M\right] + \sum_{i=2}^{M-2}\left(m\left[k_i, k_{i+1}, k_M\right] - m\left[k_1, k_i, k_{i+1}\right]\right) \\ &= -\alpha(k_M - k_1)\left\{\begin{array}{l}\left\{(k_M + k_1)(k_{M-1} - k_2) - (k_{M-1}^2 - k_2^2)\right\} \\ + (k_M + k_1)\sum_{i=2}^{M-2}(k_i - k_{i+1}) - \sum_{i=2}^{M-2}(k_i^2 - k_{i+1}^2)\end{array}\right\} = 0 \end{aligned} \tag{II.13}$$

However, direct substitution yields that conservation of the total kinetic energy due to the motion in the *x*-direction is not obeyed with the mass definition of Eq. (II.12).

Figure Captions

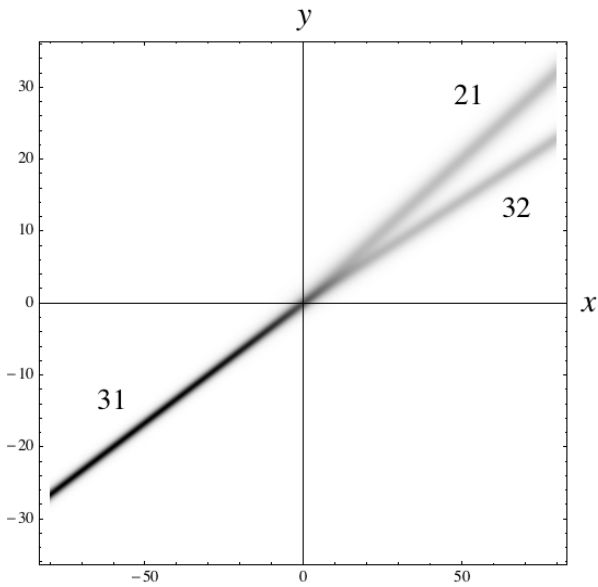

Fig. 1 *Y*-shaped KP II-soliton solution (Eqs. (5), (18)); $t = 0$; $k_1 = 1.$, $k_2 = 1.5$, $k_3 = 2.$; $\xi_1 = \xi_2 = \xi_3 = 1$.

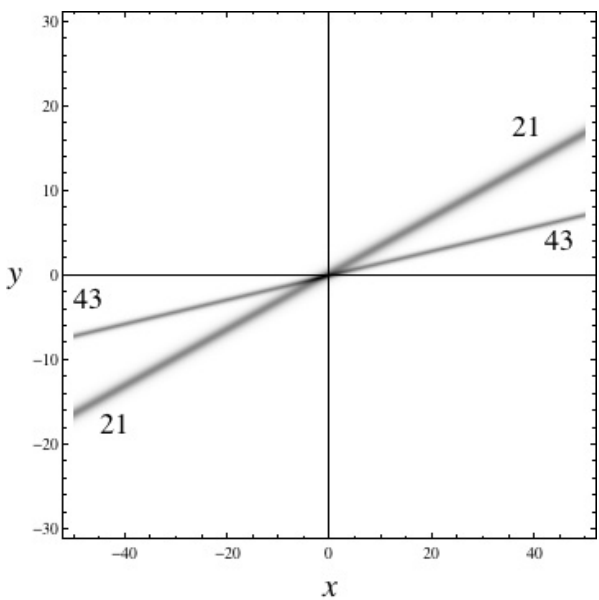

Fig. 2 *X*-shaped KP II-soliton solution (Eqs. (5), (22), $t = 0$; $k_1 = 1.$, $k_2 = 2.$, $k_3 = 3.$, $k_4 = 4.$

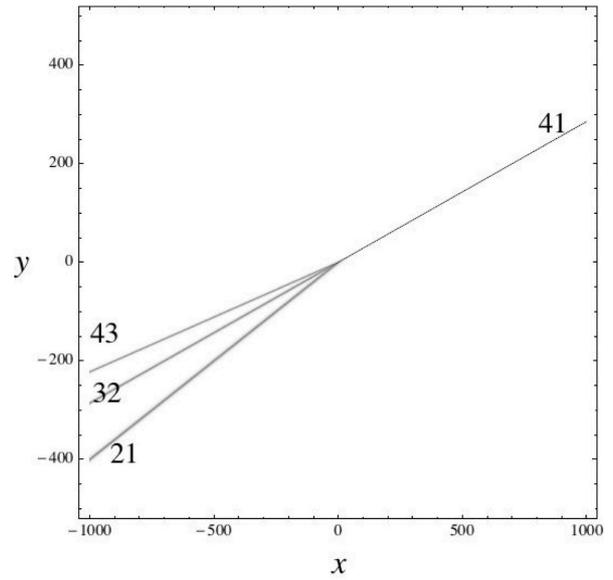

Fig. 3 KP II-soliton solution with two junctions (Eqs. (5), (25)) at $t = 0$; $k_1 = 1.$, $k_2 = 1.5$, $k_3 = 2.$, $k_4 = 2.5$; $\xi_1 = \xi_2 = \xi_3 = \xi_4 = 1$.

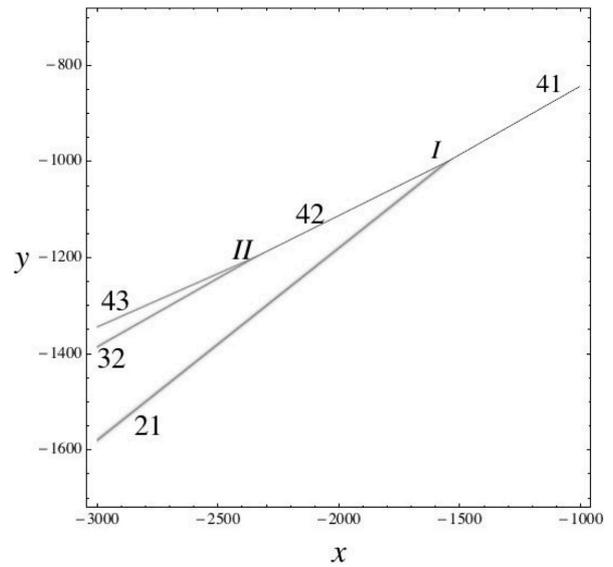

Fig. 4 KP II-soliton solution with two junctions (Eqs. (5), (25)); $t = -200$. Parameters as in Fig. 3.

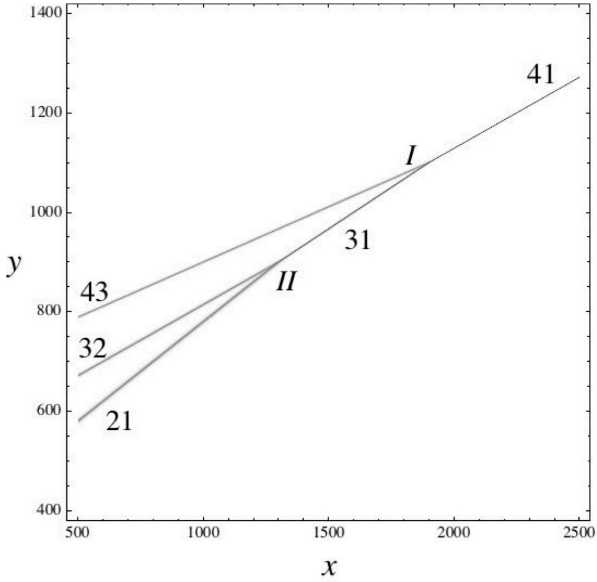

Fig. 5 KP II-soliton solution with two junctions (Eqs. (5), (25)); $t = +200$. Parameters as in Fig. 3.

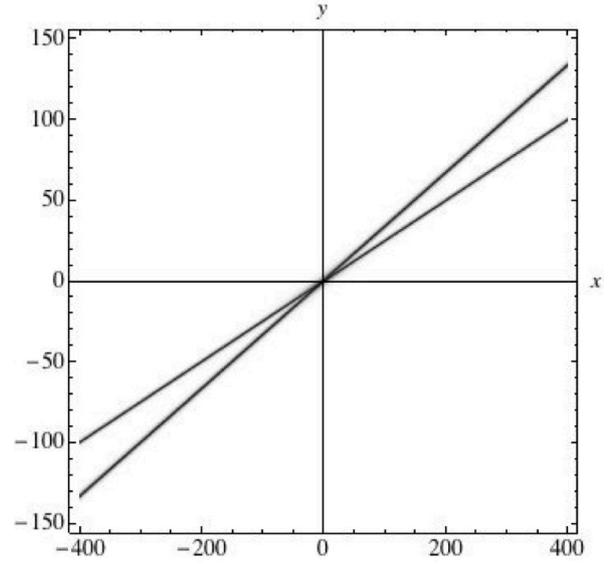

Fig. 7 KP II-soliton solution with four junctions (Eqs. (5), (28)); $t = 0$.
Parameters as in Fig. 6.

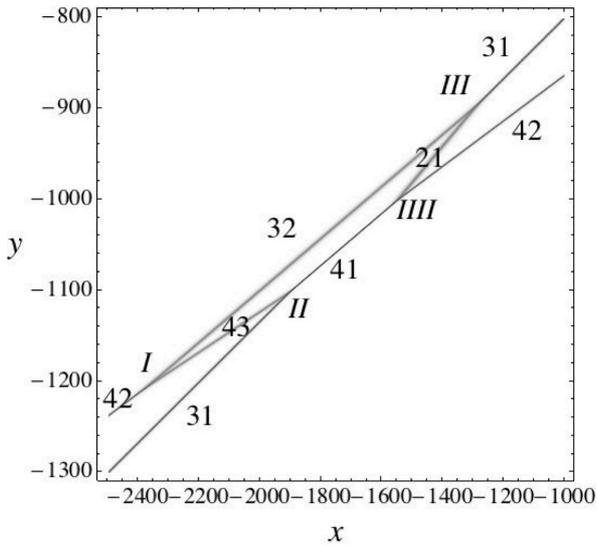

Fig. 6 KP II-soliton solution with four junctions (Eqs. (5), (28)); $t = -200$, $k_1 = 1.$, $k_2 = 1.5$, $k_3 = 2.$, $k_4 = 2.5$; $\{\xi_{12}, \xi_{13}, \xi_{14}, \xi_{23}, \xi_{24}, \xi_{34}\} = \{1,1,1,1,2,1\}$.

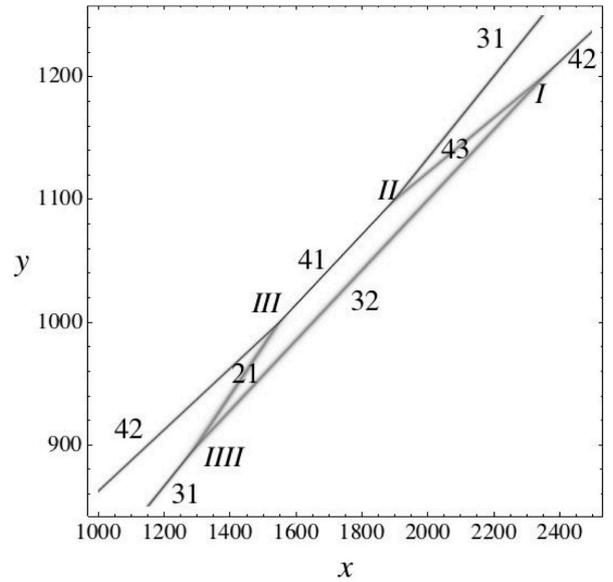

Fig. 8 KP II-soliton solution with four junctions (Eqs. (5), (28)); $t = +200$.
Parameters as in Fig. 6.

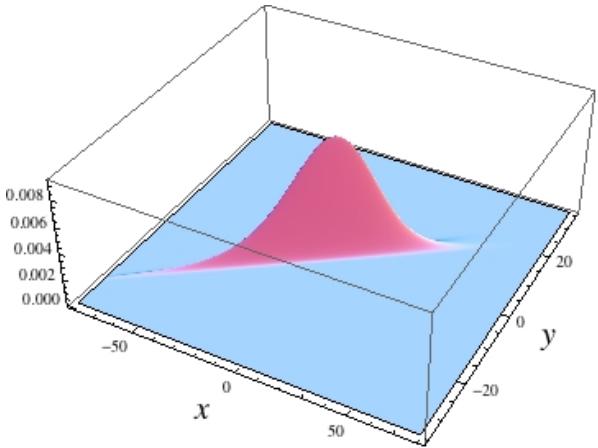

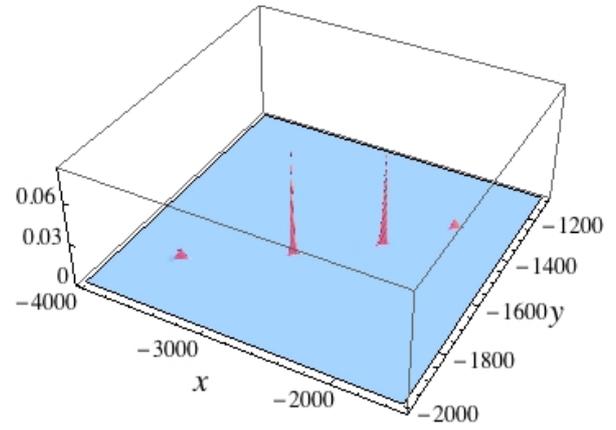

Fig. 9 Vertex map of *Y*-shaped KP II-soliton solution (Eqs. (5), (18)); $t = 0$. Parameters as in Fig. 1.

Fig. 11 Vertex map of KP II-soliton solution with four junctions (Eqs. (5), (28)); $t = -300$. Parameters as in Fig. 6.

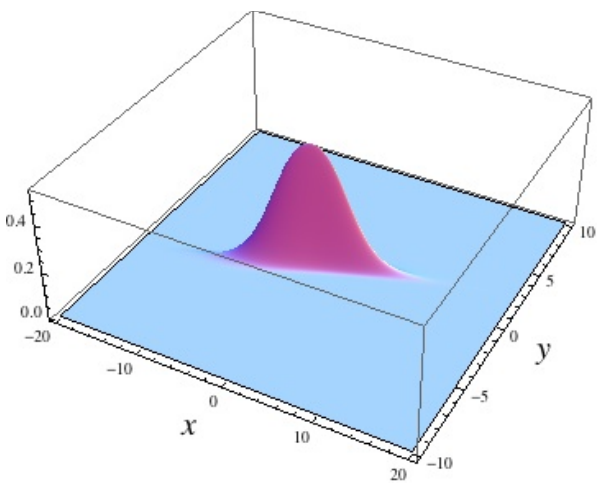

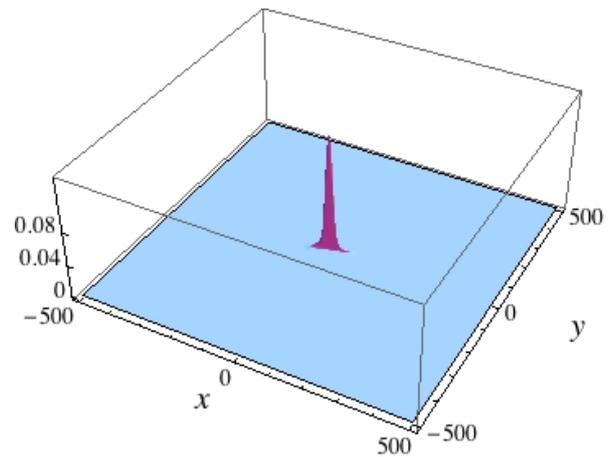

Fig. 10 Vertex map of *X*-shaped KP II-soliton solution (Eqs. (5), (23)); $t = 0$. Parameters as in Fig. 2.

Fig. 12 Vertex map of KP II-soliton solution with four junctions (Eqs. (5), (28)); $t = 0$. Parameters as in Fig. 6.

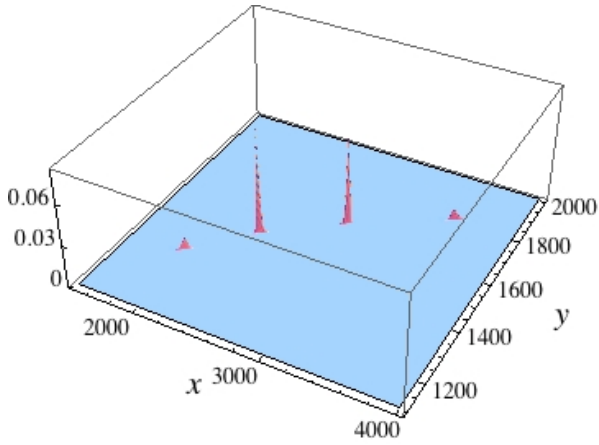

Fig. 13 Vertex map of KP II-soliton solution with four junctions (Eqs. (5), (28)); $t = +300$. Parameters as in Fig. 6.

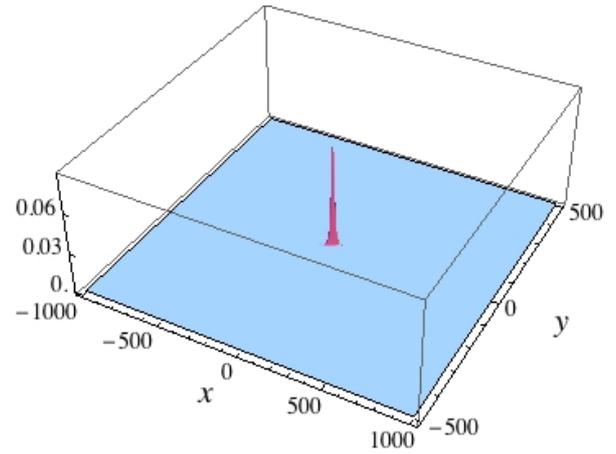

Fig. 15 Vertex map of KP II-soliton solution with two junctions (Eqs. (5), (25)); $t = 0$. Parameters as in Fig. 3.

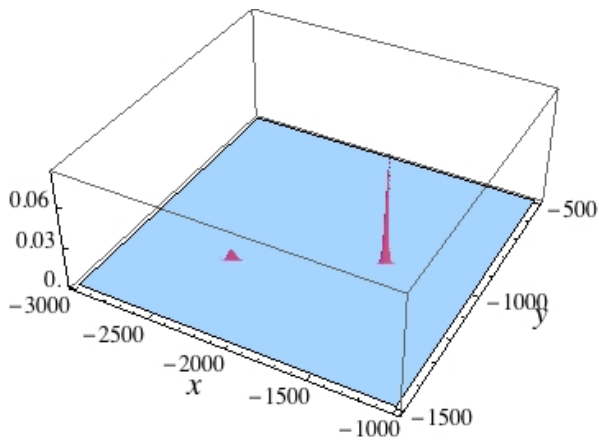

Fig. 14 Vertex map of KP II-soliton solution with two junctions (Eqs. (5), (25)); $t = -200$, Parameters as in Fig. 3.

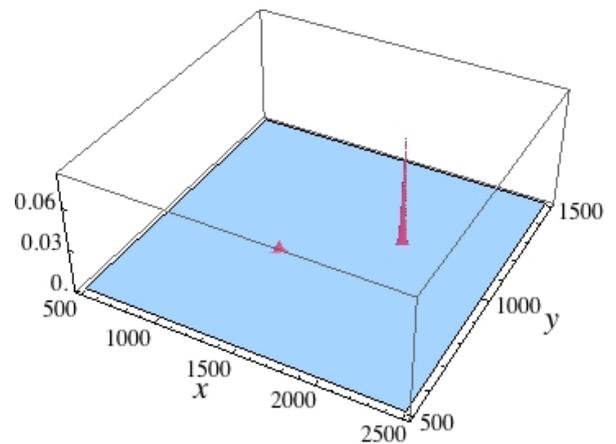

Fig. 16 Vertex map of KP II-soliton solution with two junctions (Eqs. (5), (25)); $t = +200$. Parameters as in Fig. 3.